%% file: main.tex
\documentclass[doublecol]{epl2} 
\usepackage[english]{babel}
\usepackage{amsmath,
amssymb,amsfonts,latexsym}
\usepackage{graphicx}
\usepackage{color}
\usepackage{subfigure}
\usepackage{afterpage}
\usepackage{booktabs}
\usepackage{blkarray}
\usepackage{cite}
\usepackage{ulem}
\usepackage{bm}
\usepackage{braket}
\usepackage{nicefrac}
\usepackage{etoolbox}
\usepackage{mathrsfs} 
\usepackage[protrusion=true,expansion=true]{microtype}
\usepackage{amsmath}
\usepackage[usenames,dvipsnames,table]{xcolor}
\usepackage[colorlinks,
citecolor=blue,linkcolor=red,urlcolor=blue]{hyperref}
\usepackage{amsfonts,amsmath,amssymb}
\usepackage{soul} 
\usepackage{enumitem} 
\usepackage{blindtext} 
\definecolor{Blue}{rgb}{0.00, 0.00, 1.00}
\definecolor{Red}{rgb}{1.00, 0.00, 0.00}
\definecolor{labelkey}{cmyk}{.1,.7,0.5,0}

\newcommand{\I}{{\rm i}}
\newcommand{\qqq}{\end{document}}

\newcommand{\rev}[1]{{#1}}

\include{new_commands}

\title{Domain wall melting across a defect}
\shorttitle{Domain wall melting across a defect} 
\author{Luca Capizzi \inst{1}   \and   Stefano Scopa\inst{1} \and Federico Rottoli\inst{1} \and Pasquale Calabrese\inst{1,2} }
\shortauthor{L. Capizzi \etal}
\institute{                    
  \inst{1} SISSA and INFN, via Bonomea 265, 34136 Trieste, Italy\\
 \inst{2}International Centre for Theoretical Physics (ICTP), I-34151, Trieste, Italy }
\date{\today}
\abstract{
We study the melting of a domain wall in a free-fermionic chain with a localised impurity. 
We find that the defect enhances quantum correlations in such a way that even the smallest scatterer leads to a linear growth of the entanglement entropy contrasting the logarithmic behaviour in the clean system.
Exploiting the hydrodynamic approach and the quasiparticle picture, we provide exact predictions for the evolution of the entanglement entropy for arbitrary bipartitions.
In particular, the steady production of pairs at the defect gives rise to non-local correlations among distant points. 
We also characterise the subleading logarithmic corrections, highlighting some universal features.
}
\begin{document}

\maketitle
{\bf Introduction}~--~{A single localised impurity or defect can alter the global structure of a many-body quantum system, as well known from the textbook examples of Anderson orthogonality catastrophe \cite{a-67} and the Kane-Fisher model \cite{cl-92,cl-92b}.
In the latter, it has been shown that for repulsive  interactions, the electrons are completely reflected by even the smallest scatterer, leading to a truly insulating weak link disconnecting the two halves. Conversely for attractive bulk interactions, the weak link is irrelevant, i.e., it is washed away at large scales. 
As a consequence free fermions represent the most interesting system in which the defect is marginal and there is a line of fixed points characterised by the the defect strength \cite{bddo-02,bm-06}.

In recent years, the physics of impurities in one-dimensional (1D) free-fermionic systems has been investigated a lot through the lens of entanglement. 
The marginality of the defect is reflected into a logarithmic scaling of the entanglement entropy with a prefactor that depends continuously on the defect strength \cite{p-05,ss-08,ep-12,ep-10,cmv-11c,bb-15,gm-17,cmc-22,cmc-22b,mt-21,rpr-22,kmcm-21,rs-22,cmv-12a,Klich2009}.
Overall, thanks to all these studies nowadays we have a rather complete understanding of the physics of defects in {\it equilibrium} free fermionic systems. 
The same is definitively not true when the free fermionic chain is driven {\it out of equilibrium}; in fact, in spite of several works about the non-equilibrium behaviour across one defect (see, e.g., Refs.~\cite{ep-12b,cc-13,ge-20,rcgf-22,ep-07,ce-22,Gamayun2020,Gamayun2021,DeLuca2020,Song2011,Igloi2009}), a complete understanding is still far because of the many different ways of driving a system away from equilibrium.  

In this manuscript, we prepare an initial state with a domain wall localised at the defect and we let it melt. 
Without the defect, this is a protocol that has been studied intensively in the free fermionic literature \cite{Antal1999,Antal2008,Scopa2021,Dubail2017,Allegra2016,Karevski2002,Hunyadi2004,Platini2005,Platini2007,DeLuca2013,DeLuca2014,Viti2016,rsc-22,scd-21} and it represents a case study for the application of hydrodynamics to non-equilibrium quantum systems (recently adapted also to interacting integrable models \cite{cdv-18,Collura2020} with the generalised hydrodynamics formalism \cite{cdy-16,bcdf-16}). 
Anyhow, the presence of the defect is expected to alter dramatically the evolution at a qualitative level.
The density, the currents, and other local quantities have been characterised in Ref. \cite{lsp-19} where the emergence of a local non-equilibrium stationary state (NESS) has been rigorously established.
\rev{However, little is known for the entanglement entropy, whose behaviour is affected, as any other non-local observable, by non-local correlations generated by the defect.
Some lattice results have been derived for the domain wall melting with defect in Ref.~\cite{ep-12b}, and another important step forward has been done by Fraenkel and Goldstein \cite{fg-21,fg-22} in a slightly different context}, but a general scheme to describe non-local correlations is still missing.
In this regard, it is very natural to wonder whether even a small defect alters substantially the domain wall melting. 
As we shall see, this is the case. \\
\indent{\bf The model and the quench protocol}~--~We consider a 1D chain  of free spinless fermions with $2N$ sites and with nearest-neighbour hopping with a defect of strength $\lambda$ located at the center of the system.  
The Hamiltonian is
\be\label{eq:model}
\Ha=\sum_{i,j=-N+1}^{N} h_{i,j} \ \hat{c}^\dagger_i\hat{c}_j
\ee
with
\be
h_{i,j}=-\frac{1}{2}(\delta_{i,j+1}+\delta_{i+1,j}),  \quad \forall i,j\neq 0,1
\ee
and defect
\begin{align}\label{eq:defect}
&h_{0,1}= h_{1,0}=-\frac{\lambda}{2}, 
&h_{0,0}= -h_{1,1}=\frac{1}{2}\sqrt{1-\lambda^2}.
\end{align}
Here $\hat{c}^\dagger_j$, $\hat{c}_j$ are the creation and annihilation operators of spinless fermions at site $j$, satisfying 
$\{\hat{c}^\dagger_j,\hat{c}_i\}=\delta_{ij}$. For $\lambda=1$ the Hamiltonian \eqref{eq:model} reduces to a standard hopping model. The system is initially prepared in the product state
\be\label{eq:initial-state}
\ket{\psi_0}=\bigotimes_{j=-N+1}^{0} \ket{1}_j \bigotimes_{j=1}^{N} \ket{0}_j,
\ee
where $\ket{\zeta=0,1}_j$ are the eigenstates of the fermionic number operator $\hat{c}^\dagger_j\hat{c}_j$ with eigenvalues $\zeta=0,1$. In the spin language, the initial state \eqref{eq:initial-state} corresponds to the domain wall $\ket{\psi_0}=\ket{\dots\uparrow\uparrow\downarrow\downarrow\dots}$. \rev{Thereafter, our discussions can be extended to the XX spin chain with defect \cite{ep-10},  obtained from the model \eqref{eq:model} through a Jordan-Wigner transformation.} For $t>0$, the state \eqref{eq:initial-state} is unitarily evolved with Hamiltonian \eqref{eq:model}, $\ket{\psi_t}=e^{-\I t\Ha}\ket{\psi_0}$.\\

The structure of the defect \eqref{eq:defect} does not spoil the exact solvability of the free fermionic model, which has spectrum $\omega_q=-\cos(k_q)$, $k_q=\pi q/(2N)$ ($q=1,\dots,2N$), for any value of $\lambda\in(0,1]$ \cite{ce-22,ep-12}. Moreover, the eigenstates of $\Ha$ can be related to the eigenstates $\Psi_{q,2N}(j)=\sin(k_q j)/\sqrt{N}$ of \eqref{eq:model} in the absence of defect ($\lambda=1$) as \cite{ce-22}
\be
\Psi^\text{def}_{q,2N(j)}=\Theta(-j)\alpha^+_q(\lambda)\Psi_{q,2N} 
+ \Theta(j)\alpha^-_q(\lambda) \Psi_{q,2N},\ee
with $\Theta(j)$ the Heaviside step function and coefficients $\alpha^\pm_q(\lambda)=[1\pm(-1)^q\sqrt{1-\lambda^2}]^{1/2}$. For $N\to\infty$, this eigenproblem reduces to a scattering of plane waves across a localised defect, i.e.,
\be
\Psi^\text{def}_{k,\infty}(j)\propto\Theta(j) \lambda \ e^{\I k j} + \\
\Theta(-j)(\sqrt{1-\lambda^2} \ e^{-\I k j} + e^{\I kj})
\ee
with transmission probability ${T}(\lambda)\equiv \lambda^2$ and reflection probability ${R}(\lambda)\equiv 1-\lambda^2$. These parameters do not depend on the momentum $k$ of the scattered particle and so the defect \eqref{eq:defect} is also known as {\it conformal defect} \cite{ep-12,ep-12b,ep-10}. 

\begin{figure}[t]
\centering
\includegraphics[width=.65\columnwidth]{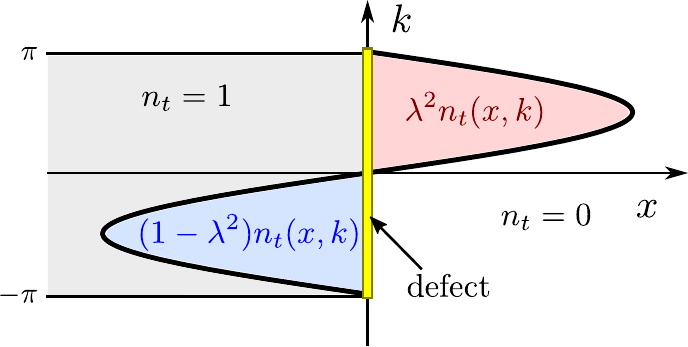}
\caption{Illustration of the evolution of the Fermi occupation function $n^{(\lambda)}_t(x,k)$ in the presence of the defect.
The light-grey area is the initial occupation \eqref{n0}, while the colored regions correspond to the time-evolved one.
}\label{fig:DW}
\end{figure}
\indent
{\bf Hydrodynamic limit}~--~
Exact asymptotic results for the charges profiles can be obtained in the hydrodynamic limit $N\to\infty$, $j\to\infty$, $t\to \infty$ at fixed $j/t$.
The lattice index $j$ and the quantised momenta $k_q$ are replaced by continuous variables for the position $x=j a\in\mathbb{R}$ 
($a$ is the lattice spacing) and for the momenta $-\pi\leq k \leq \pi$. 
In such scaling limit, the essential information on the initial state \eqref{eq:initial-state} is retained by the local fermionic occupation
\be
n_0(x,k)=\begin{cases} 1, \quad \text{if $x\leq 0$ and $-\pi\leq k \leq \pi$};\\[3pt]
0, \quad\text{otherwise}.
\end{cases}
\label{n0}
\ee
In the absence of defect ($\lambda=1$), the evolution of the occupation function is given by the Euler equation
\be\label{eq:GHD}
(\de_t +\sin k \ \de_x) n_t(x,k)=0,
\ee
which can be simply solved as 
\be
n_t(x,k)=n_0(x-t\sin k , k),
\label{nth}
\ee
i.e. it stays equal to $0$ and $1$ and only the Fermi contour separating the two values evolves. 
Intuitively, this solution encodes the fact that each non-interacting particle moves along the ballistic trajectory with constant velocity $v(k)=\sin k$. 
Instead, for $\lambda\neq 1$, a particle of momentum $k>0$ traveling from $x<0$ is scattered by the defect in such a  way that it is reflected with probability $R(\lambda)$ and transmitted with probability $T(\lambda)$. 
\begin{figure}
\includegraphics[width=.9\columnwidth]{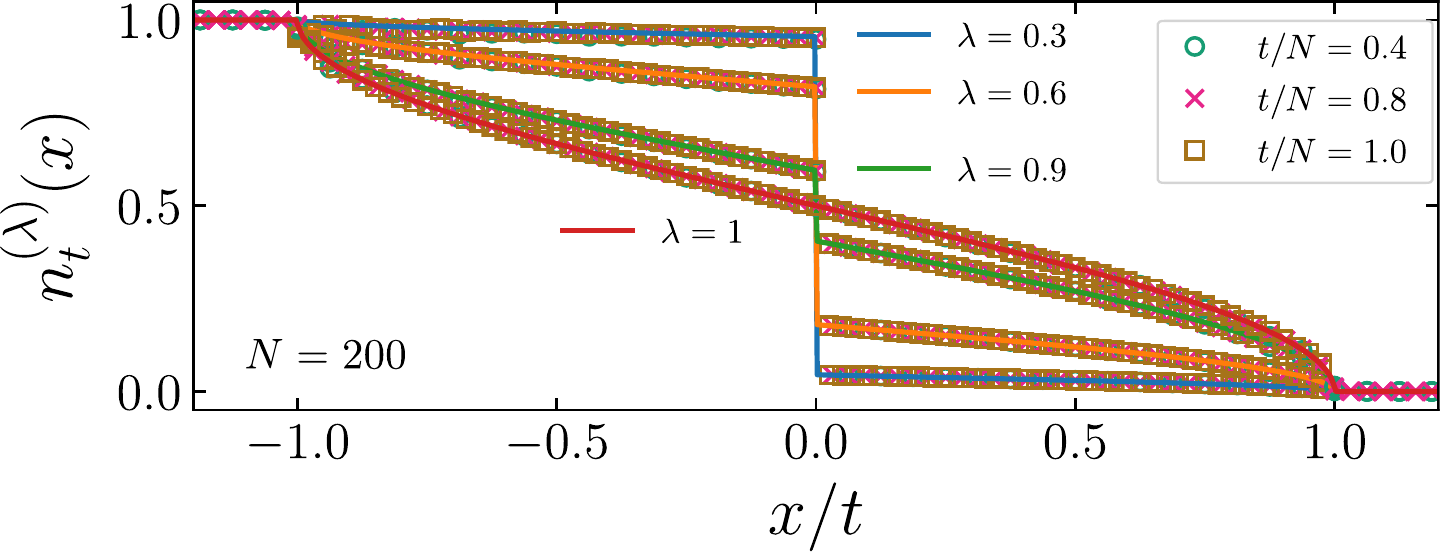}
\caption{Fermionic density as function of  $x/t$ for different values of $\lambda$ and $t$. Symbols show the numerical data obtained with exact lattice calculations with $400$ sites while the full lines are given by Eq.~\eqref{eq:density}.}\label{fig:density}
\end{figure}
Accordingly, the time-evolved occupation function in the presence of the defect takes the form \cite{lsp-19}
\begin{multline}
    \label{eq:n_t}
n^{(\lambda)}_t(x,k)=\lambda^2 \Theta(x) n_t(x,k)+\\
 + \Theta(-x)\left[ (1-\lambda^2)n_t(-x,-k) + n_t(x,k)\right],
\end{multline}
as illustrated in Fig.~\ref{fig:DW}. In our notations 
$n_t^{(\lambda\equiv1)}= n_t$ in Eq.~\eqref{nth}. 
The occupation function \eqref{eq:n_t} gives us access to the asymptotic profiles of conserved charges as elementary integrals over the modes $k$, properly weighted with the single-particle eigenvalue of the associated charge~\cite{Antal2008,Antal1999}. 
For instance, the particle density profile for $0< x\leq t$ is 
\be\label{eq:density}
n^{(\lambda)}_t(x)=\int_{-\pi}^\pi \frac{\dd k \ n^{(\lambda)}_t(0<x\leq t,k)}{2\pi}=\lambda^2 \frac{\arccos(x/t)}{\pi}.
\ee
For $-t\leq x<0$, the profile is obtained via particle-hole symmetry and reads $n^{(\lambda)}_t(x)=1-\lambda^2 \arccos(|x|/t)/\pi$. Outside the correlated region, i.e., for $|x|>t$, the systems keeps its initial configuration with constant density $n^{(\lambda)}_t=1$ ($n^{(\lambda)}_t=0$) on its left (right) part. In Fig.~\ref{fig:density}, we compare numerical results obtained from exact lattice calculations with the hydrodynamic result \eqref{eq:density}.

The crucial observation now is that the local occupation function \eqref{eq:n_t} for $\lambda\neq1$ assumes values which are different from $0$ and $1$. Consequently, the local entropy is non vanishing, resulting in an extensive entanglement. Physically this phenomenology roots back to the correlation between reflected and transmitted modes generated at the defect.

{\bf Entanglement dynamics}~--~
We now  move to our main goal which is characterising the entanglement dynamics. Specifically, we focus on a bipartition of the system $A\cup B$ with a reduced density matrix $\hat\rho_t(A)={\rm tr}_B\ket{\psi_t}\bra{\psi_t}$. 
The $n$-R\'enyi entropy is
\be\label{eq:Renyi}
S_n(A,t)=\frac{1}{1-n}\log {\rm tr}[\hat\rho_t(A)^n]
\ee
that provides the entanglement entropy in the limit $n\to1$, i.e. $
S_1(A,t)=-{\rm tr}[\hat\rho_t(A)\log\hat\rho_t(A)]$.

\begin{figure}
\centering
\includegraphics[width=\columnwidth]{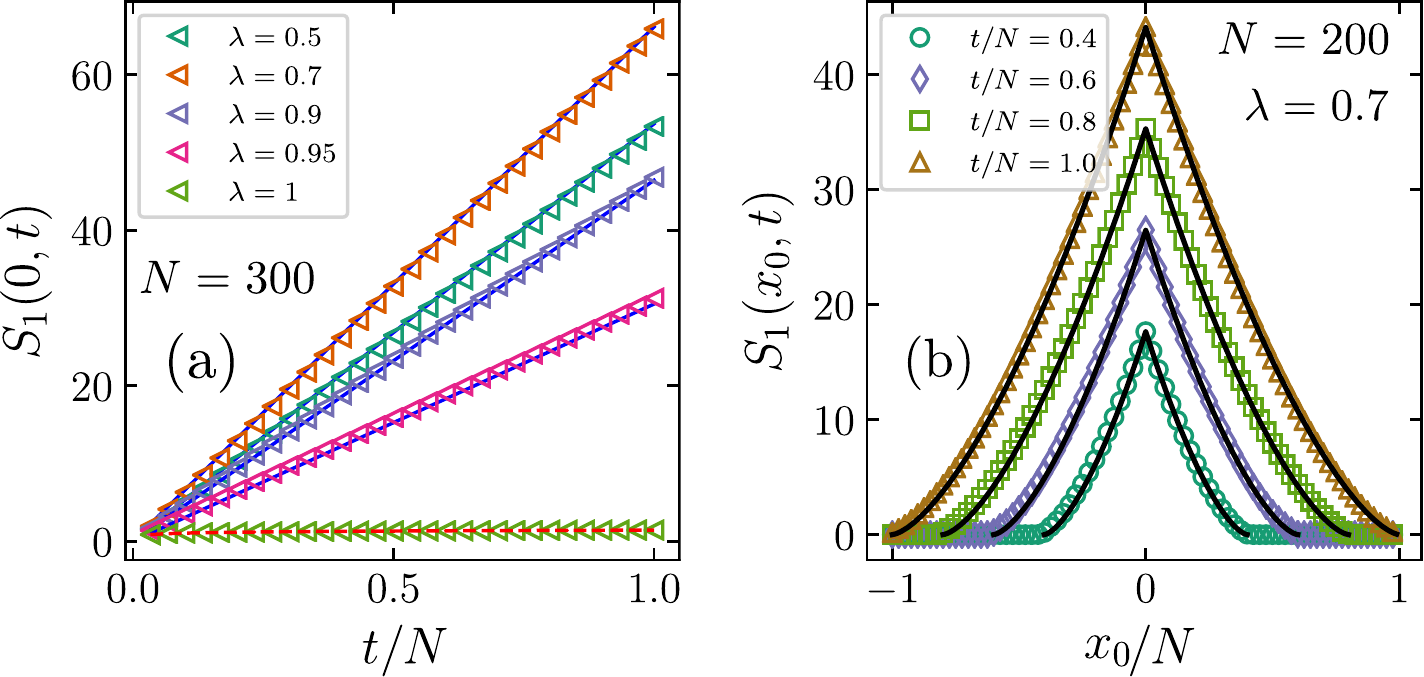}
\caption{(a)~--~Half-system entanglement of $A=[-\infty,0]$ for different values of $\lambda$ as function of time. Symbols show the numerical data while the full lines (for $\lambda\neq 1$) are given by Eq.~\eqref{eq:renyi-QPC-half-sys}. At $\lambda=1$, the half-system entanglement entropy is $S_1=1/6\log(t)+\text{const}$ (dashed line)~\cite{Dubail2017}. (b)~--~Entanglement profiles for $A=[-\infty,x_0]$ plotted as function of $x_0$ at different times and fixed $\lambda=0.7$. Symbols show the numerical data while the full lines are given by Eq.~\eqref{eq:renyi-QPC}.
}
\label{fig:half-sys}
\end{figure}

An ab-initio description of the entanglement dynamics is very demanding even in the absence of defect, due to the non-equilibrium and non-homogeneous character of the quench problem under analysis. However, the asymptotic behaviour of entanglement can be determined with hydrodynamic arguments as follows.
First of all, we recall the definition of 
the local Yang-Yang R\'enyi entropy \cite{Bertini2018,fg-22,bka-22,ac-17}
\be
s_n(x,t)\equiv \hspace{-1mm}
\frac{1}{1-n} \int_{-\pi}^\pi \hspace{-.5mm}
\frac{\dd k}{2\pi} \log\Big[n_t^{(\lambda)}(x,k)^n 
 + (1-n_t^{(\lambda)}(x,k))^n\Big],
\ee
which is non-zero because $n_t^{(\lambda)}\neq0,1$. 
From an entanglement perspective, this entropy measures the correlations between the particles at position $x$ with the ones at $-x$ generated by the scattering at the defect, according to the quasiparticle picture \cite{Calabrese2005,Alba2017}. 
The total entanglement entropy of a region $A$ which is entirely to the right or to the left of the defect (say $A=[-\infty,x_0]$) is then given by (see also~\cite{Gamayun2020,cdy-16})
\be
\label{eq:renyi-QPC}
S_n(x_0,t)=\int_{A} \dd x s_n(x,t) =
\frac{{\cal N}_t(A)}{1-n}\log\left[\lambda^{2n}+(1-\lambda^2)^n\right],
\ee
with ${\cal N}_t(A)$ being the total number of entangled particles in the region $A$ at time $t$. 
For example for $A=[-\infty,x_0]$ with $x_0<0$, we have
\be
{\cal N}_t([-\infty,x_0])=\frac{t}{\pi}\left(\sqrt{1-\frac{x_0^2}{t^2}}-\frac{x_0}{t}\arccos\frac{x_0}{t}\right).
\ee
By setting $x_0=0$, Eq.~\eqref{eq:renyi-QPC} predicts a linear growth of entanglement \rev{(see also Ref.~\cite{ep-12b})}
\be\label{eq:renyi-QPC-half-sys}
S_n(0,t)= 
\frac{t}{\pi(\rev{1-n})} \log\left[\lambda^{2n}+(1-\lambda^2)^n\right].
\ee
In the absence of defect, previous studies highlighted a half-system entanglement growth $S_n(0,t)\sim (n+1)/12n \log(t)$, arising from subleading contributions \cite{Ares2022,scd-21,Dubail2017}. Interestingly, the entanglement transition from logarithmic to linear law is observed even for values of $\lambda$ very close to unit, see Fig.~\ref{fig:half-sys}-(a) \rev{for a comparison with exact lattice calculations. We refer to e.g.~Refs.~\cite{Scopa2021, Ares2022} for details on the numerical implementation.}

Eq.~\eqref{eq:renyi-QPC} fails to capture the behaviour of entanglement for a subsystem straddling the defect because it counts also for the pairs of entangled particles which are both in $A$, but on different sides of the defect. 
Such over-counting is however easily cured within the quasiparticle picture~\cite{Calabrese2005,Alba2017}.
First, for the case $A=[-\infty,x_0]$ with $x_0>0$, using particle-hole symmetry 
and $S_n(A,t)=S_n(\bar{A}, t)$, we have 
$S_n(x_0,t)= S_n(-x_0,t)$ where the rhs is in Eq.~\eqref{eq:renyi-QPC}.  
The validity of Eq.~\eqref{eq:renyi-QPC} is tested against exact lattice calculations in Fig.~\ref{fig:half-sys}-(b).

\begin{figure}[t]
\centering
\includegraphics[width=0.35\textwidth]{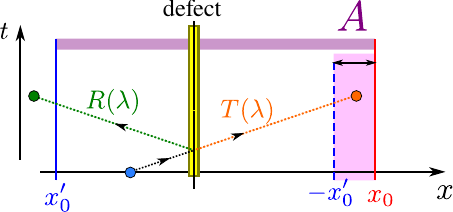}
\caption{Illustration of the quasiparticle picture. The partial reflection $R(\lambda)$ and transmission $T(\lambda)$ at the defect causes entanglement between symmetric points. When computing the entanglement entropy of a sybsystem straddling the defect, the Yang-Yang entropy overcounts the quasiparticles: the correct counting is given by the shaded pink area in the figure.}
\label{QPP}
\end{figure}
\begin{figure}
\centering
\includegraphics[width=\columnwidth]{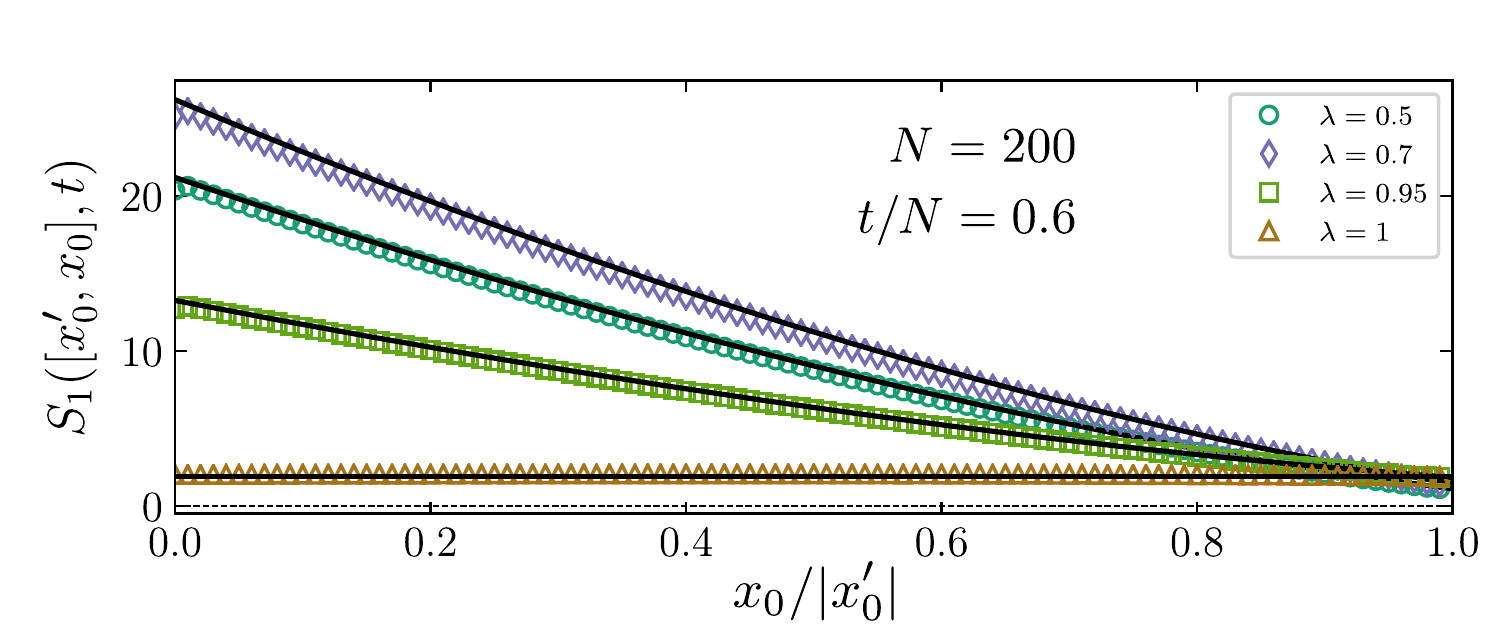}
\caption{Entanglement entropy for $A=[x_0',x_0]$ at fixed time $t/N=0.6$ and for different values of $\lambda$, plotted as function of the left endpoint $x_0$. Symbols show the numerical data while the full lines are given by Eq.~\eqref{eq:renyi-QPC-twoblocks}, up to a fitted additive constant. 
}\label{fig:interval}
\end{figure}
\setlength{\belowcaptionskip}{0pt}
For subsystems $A=[x_0',x_0]$ consisting of an interval straddling the defect (i.e., $x'_0<0$ and $x_0>0$), we can simply correct the over-counting of Eq.~\eqref{eq:renyi-QPC} by subtracting the doubly-counted particles (see Fig.~\ref{QPP} for an illustration), resulting finally in
\be\label{eq:renyi-QPC-twoblocks}
S_n([x_0',x_0],t)=\left|S_n(|x_0|,t)-S_n(|x_0'|,t)\right|.
\ee
In Fig.~\ref{fig:interval}, numerical results for the lattice model are compared with the hydrodynamic prediction in Eq.~\eqref{eq:renyi-QPC-twoblocks}, showing an excellent agreement.

Our results also show the appearance of long-range entanglement in the sense of Ref. \cite{fg-22}. 
Indeed, if we compute the mutual information $I_{A_1:A_2}\equiv S_{A_1}+S_{A_2}-S_{A_1\cup A_2}$ between $A_1=[-\infty,-x_0]$
and $A_2=[x_0,\infty]$ we have
\be
I_{A_1:A_2}= 2 S_n([x_0,\infty],t)\,
\ee
because $S_{A_1\cup A_2}=0$, (up to subleading terms, see also the next section). The same remains true if $A_1$ and $A_2$ are two symmetric finite intervals. Such large (actually extensive for large $t$) mutual information is due to the constant presence of shared pairs 
between symmetric intervals, exactly as in Ref. \cite{fg-22}. 

{\bf Subleading behaviour}~--~When the   subsystem $A$ is placed symmetrically across the defect, i.e., $A=[-x_0,x_0]$, the entanglement resulting from correlated pairs of particles in Eq.~\eqref{eq:renyi-QPC-twoblocks} vanishes.
This is clearly due to the fact that entangled pairs have symmetric positions and so they are either both in $A$ or in the complement.  
As a consequence, the behaviour of entanglement is entirely due to subleading contributions associated with quantum fluctuations, which cannot be determined with a semiclassical approach.
For the homogeneous Hamiltonian ($\lambda=1$), a useful way to incorporate quantum fluctuations in our description is established by quantum generalised hydrodynamics \cite{Scopa2021,scd-21,Ares2022,Collura2020,Ruggiero2020,Scopa-Horvath22,Ruggiero2019,rcdd-22}. According to this theory, the relevant contribution to the entanglement in zero-entropic states is given by linear quantum fluctuations $\delta\hat{n}_t(x)$ at the edges of $n_t(x,k)$, corresponding to the formation of particle-hole pairs near the local Fermi points.

\begin{figure}
\centering
\includegraphics[width=\columnwidth]{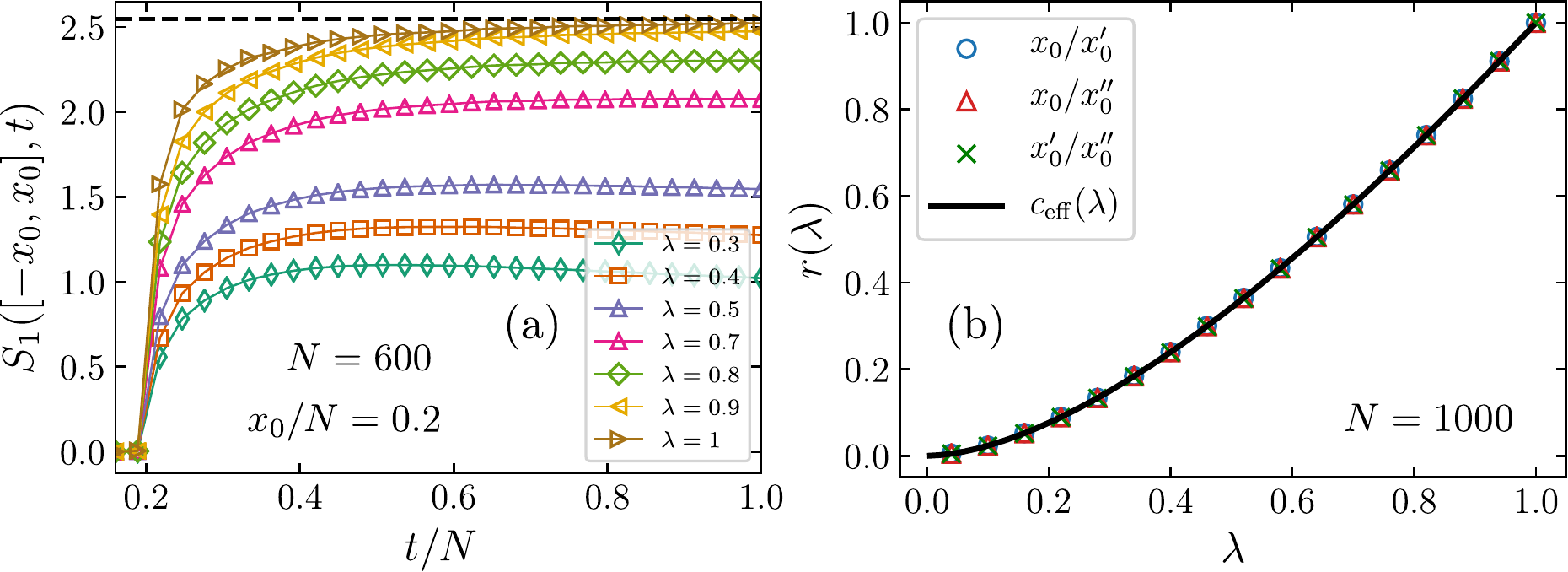}
\caption{(a)~--~Numerical results for \rev{the entanglement of the symmetric interval} $A=[-x_0,x_0]$ for different values of $\lambda$ as function of time. The dashed horizontal line mark the plateau $S_1=1/3\log(x_0)+2\kappa_1$ for $\lambda=1$. (b)~--~Plot of the ratios $r(\lambda)$ in Eq.~\eqref{eq:ratios} as function of $\lambda$ and different values of the interval size $x_0=20$, $x_0'=40$, $x_0''=60$. The full line shows the behaviour of $c_\text{eff}(\lambda)$, given by Eq.~\eqref{eq:ceff}.}\label{fig:saturation}
\end{figure}

The time evolution of the entanglement entropy is a straightforward but tedious adaptation of the calculations reported, e.g., in Ref. \cite{scd-21} for a slightly different situation.
We only report here the final result (which indeed coincides with the one in Refs. \cite{scd-21,Eisler2021})
\be\label{eq:doubleDW}
S_n([-x_0,x_0],t)=\frac{n+1}{12n}\log\left[x_0^2(1-{x_0^2}/{t^2})^3\right]+2\kappa_n,
\ee
with $\kappa_n$ a known non-universal amplitude \cite{Jin2004,Calabrese2010},  for instance $\kappa_1\simeq 0.4785$. 
For $t\gg x_0$, Eq.~\eqref{eq:doubleDW} predicts a saturation of the half-system entanglement to the value $S_1([-x_0,x_0],t\gg x_0)\approx 1/3\log(x_0)+2\kappa_1$. 

Numerical exact calculations for the lattice model reveal a similar behaviour for the half-system entanglement even in the presence of the defect $\lambda\in(0,1)$, see Fig.~\ref{fig:saturation}-(a). 
We expect that the large-time plateaus in the figure scale like $\log(x_0)$ for large $x_0$, i.e., we expect 
\be \label{eq:lambdasaturation}
S_1([-x_0,x_0],\infty)\sim \frac{c_\text{eff}(\lambda)}{3}\log(x_0)+\gamma(\lambda).
\ee
To have an unbiased estimate of $c_\text{eff}(\lambda)$ we proceed as following. 
We first consider the difference of plateaus reached at fixed $\lambda$ for different sizes of $A$, i.e.,
\be
\Delta S^{(\lambda)}_1(x_0,x_0')\equiv 
S_1([-x_0,x_0],\infty)-S_1([-x_0',x_0'],\infty),
\ee
(where by ``infinite time'' we just means confidently within the plateau);
then we take the ratio
\be\label{eq:ratios}
r(\lambda)=\Delta S^{(\lambda)}_1(x_0,x_0')/\Delta S^{(1)}_1(x_0,x_0'),
\ee
that for large $x_0, x_0'$ converges to 
$c_\text{eff}(\lambda)$ by construction.
The perfect collapse in Fig.~\ref{fig:saturation}-(b) of the ratios $r(\lambda)$ for different pairs $x_0, x_0'$ confirms the conjectured behaviour of Eq.~\eqref{eq:lambdasaturation}.
Moreover, the resulting factor $c_\text{eff}(\lambda)$ is numerically consistent with
the effective central charge appearing in the ground-state entanglement of free fermions with defects \cite{ep-10} given by
\begin{multline}
\label{eq:ceff}
c_\text{eff}(\lambda)= -\frac{6}{\pi^2} \bigg\{(1+\lambda)\mathrm{Li_2}(-\lambda) + (1-\lambda) \mathrm{Li_2}(\lambda) \\
 +\Big[ (1+\lambda) \log (1+\lambda) + (1-\lambda) \log (1-\lambda) \Big] \log \lambda \bigg\} \, ,
\end{multline}
that satisfies $c_\text{eff}(0)=0$ and $c_\text{eff}(1)=1$. We believe that, being these logarithmic \rev{contributions}  related to zero-point fluctuations, it should be possible to map explicitly the equilibrium entanglement to the non-equilibrium one. 
However, this goes beyond the scope of this work.\\
\indent
{\bf Summary and conclusions}~--~
We studied the time evolution of the entanglement entropy in a domain wall melting across a conformal defect. 
We showed that the pure-system logarithmic growth in time of the entanglement entropy is turned, by the smallest defect, into a linear one with an extensive stationary value corresponding to a non-vanishing thermodynamic Yang-Yang entropy. 
Furthermore we showed that there are extensive long-range correlations between sites which are mirror images of each other with respect to the defect.
None of these effects has an equilibrium counterpart. 
We also characterised the subleading logarithmic contributions, which are not captured by the quasiparticle picture. 

A natural extension of this work could be the study of the dynamics in the presence of multiple defects. In that case, we expect a richer pattern of long-range correlations arising from multiple scattering across the defects, e.g. along the lines of Ref. \cite{fg-22}. However, we still do not know how to deal systematically with those effects and how to incorporate them in a quasiparticle picture for the entanglement.

{\it Acknowledgments}. LC is grateful to Viktor Eisler for useful discussions. The authors acknowledge support from ERC under Consolidator grant
number 771536 (NEMO).

\input{bibliography.tex}
\end{document}

%% file: new_commands.tex
\newcommand{\be}{\begin{equation}}
\newcommand{\ee}{\end{equation}}

\newcommand{\beA}{\begin{align}}
\newcommand{\eeA}{\end{align}}

\newcommand{\bV}{\begin{bmatrix}}
\newcommand{\eV}{\end{bmatrix}}

\newcommand{\dd}{\mathrm{d}}
\newcommand{\de}{\partial}

\newcommand{\Ha}{\hat{H}} 

